\documentclass[10 pt,final,journal,letterpaper,oneside,twocolumn]{IEEEtran}
\usepackage{multicol}   
\usepackage{multirow}
\usepackage[pdftex]{graphicx}
\usepackage[ruled]{algorithm2e}
\usepackage{algorithmic}
\usepackage[small]{caption}
\usepackage{float}
\usepackage{amssymb,amsmath}
\usepackage{graphicx}
\usepackage{subfigure}
\usepackage{xspace}
\usepackage{amsthm}
\usepackage{fancyhdr}

\usepackage{caption,setspace}
\usepackage{amssymb,amsmath}
\usepackage{textcomp} 
\usepackage{epstopdf}
\usepackage{bbding}
\usepackage{cite} 
\usepackage{color}
\usepackage{pifont}

\begin{document}
%
\title{\huge Reconfigurable Intelligent Surfaces for 6G Non-Terrestrial Networks: Assisting Connectivity from the Sky}

\author{Wali Ullah Khan, Asad Mahmood, Chandan Kumar Sheemar, Eva Lagunas, \textit{Senior Member, IEEE},\\ Symeon Chatzinotas, \textit{Fellow, IEEE}, Bj\"orn Ottersten, \textit{Fellow, IEEE} \thanks{This work has been partially supported by the Luxembourg National Research Fund (FNR) under the project MegaLEO (C20/IS/14767486) and the project RISOTTI (C20/IS/14773976).  For the purpose of open access, the authors have applied a Creative Commons Attribution 4.0 International (CC BY 4.0) license to any Author Accepted Manuscript version arising from this submission.

Authors are with the Interdisciplinary Center for Security, Reliability and Trust (SnT), University of Luxembourg, 1855 Luxembourg City, Luxembourg, \{waliullah.khan, asad.mahmood, chandankumar.sheemar, eva.lagunas, symeon.chatzinotas, bjorn.ottersten\}@uni.lu}}%

\markboth{Accepted for Publication in IEEE Internet of Things Magazine}%
{Shell \MakeLowercase{\textit{et al.}}: Bare Demo of IEEEtran.cls for IEEE Journals} 
\maketitle

\begin{abstract}
Sixth-generation (6G) non-terrestrial networks (NTNs) are advanced wireless communication systems that operate beyond traditional terrestrial networks. These networks utilize various technologies and platforms to provide flexible, enhanced connectivity and coverage. When operating at high frequency, ground user terminals require low-directional antennas, which experience poor link budgets from satellites and thus drive the quest for novel solutions. Reconfigurable Intelligent Surfaces (RISs) have recently emerged as a promising technology for 6G and beyond cellular systems. This paper studies the potential of RIS-integrated NTNs to revolutionize next-generation connectivity. First, it discusses the fundamentals of RIS technology. Secondly, it delves into reporting the recent advances in RIS-integrated NTNs. Subsequently, it presents a novel framework based on the current state-of-the-art for IRS-integrated NTNs with classical single connected diagonal RIS and fully connected beyond diagonal RIS architectures. Finally, the paper highlights open challenges and future research directions to revolutionize the realm of RIS-integrated NTNs. 
\end{abstract}  


\IEEEpeerreviewmaketitle

\section{Introduction}
Sixth-generation non-terrestrial networks (NTNs) provide a visionary approach to wireless communications that extends beyond traditional terrestrial infrastructure. These networks encompass satellite systems, high-altitude platforms (HAPs), and unmanned aerial vehicles (UAVs), aiming to provide global coverage, ubiquitous connectivity, and support for various applications, from remote sensing to Internet of Things (IoT) services. The primary purpose of NTNs is to leverage the unique advantages of non-terrestrial platforms to address challenges faced by conventional terrestrial networks, making them a pivotal component in next-generation communications systems. The NTNs consist of multiple layers, including Geostationary (GEO), Low Earth Orbit (LEO), Stratosphere and Low Atmosphere. In spite of presenting favorable benefits, compared to terrestrial networks, NTNs can suffer several challenges, such as signal coverage and capacity in harsh environments, propagation losses in the atmosphere and space, dynamic channel conditions and high energy consumption, spectrum sharing with terrestrial networks, and security issues.
\begin{figure*}[!t]
\centering
\includegraphics[width=0.8\textwidth]{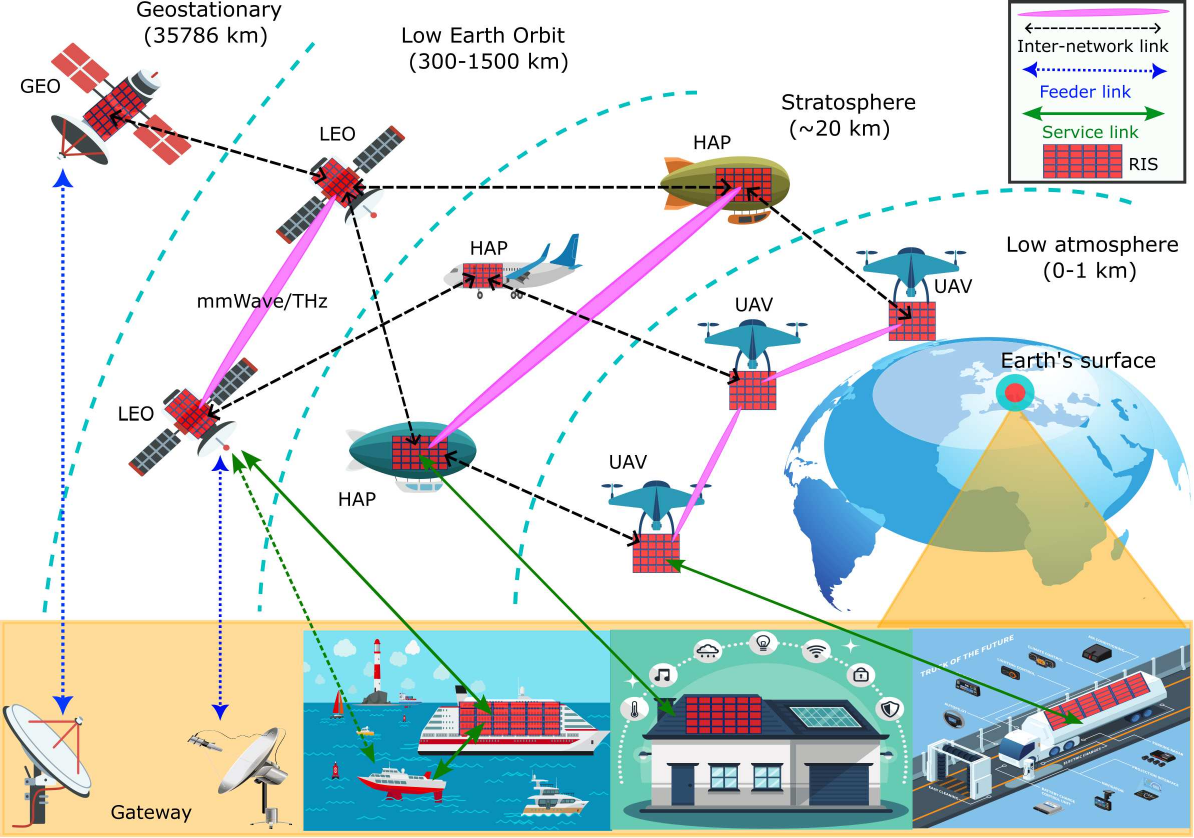}
\caption{RIS-integrated multilayered NTNs.}
\label{IoTmagN}
\end{figure*}

Reconfigurable Intelligent Surfaces (RISs) have emerged as a transformative technology in wireless communications. They consist of a large number of meta-elements that are capable of manipulating signals' phase, amplitude, and polarization. More specifically, RIS can intelligently control signal propagation by reflecting, refracting, and focusing signals toward specific locations, effectively enhancing signal strength, coverage, and link quality. Their passive nature and adaptability make them a compelling choice for improving communication performance and energy efficiency. Depending on the architecture, RIS can be categorized into three categories: 1) \emph{single connected}, 2) \emph{fully connected}, and 3) \emph{group connected}.  In the single connected case, elements are not interconnected, and the phase-shift response is diagonal. In the fully connected case, elements are fully interconnected, and the phase shift response is the full matrix. In the group connected case, elements are divided into groups, with each group being fully interconnected, and the phase-shift response results are block diagonal. Note that the single connected architecture is also known as classical diagonal RIS, while the fully connected and group connected architectures are called beyond diagonal RIS.

\begin{figure}[!t]
\centering
\includegraphics[width=0.45\textwidth]{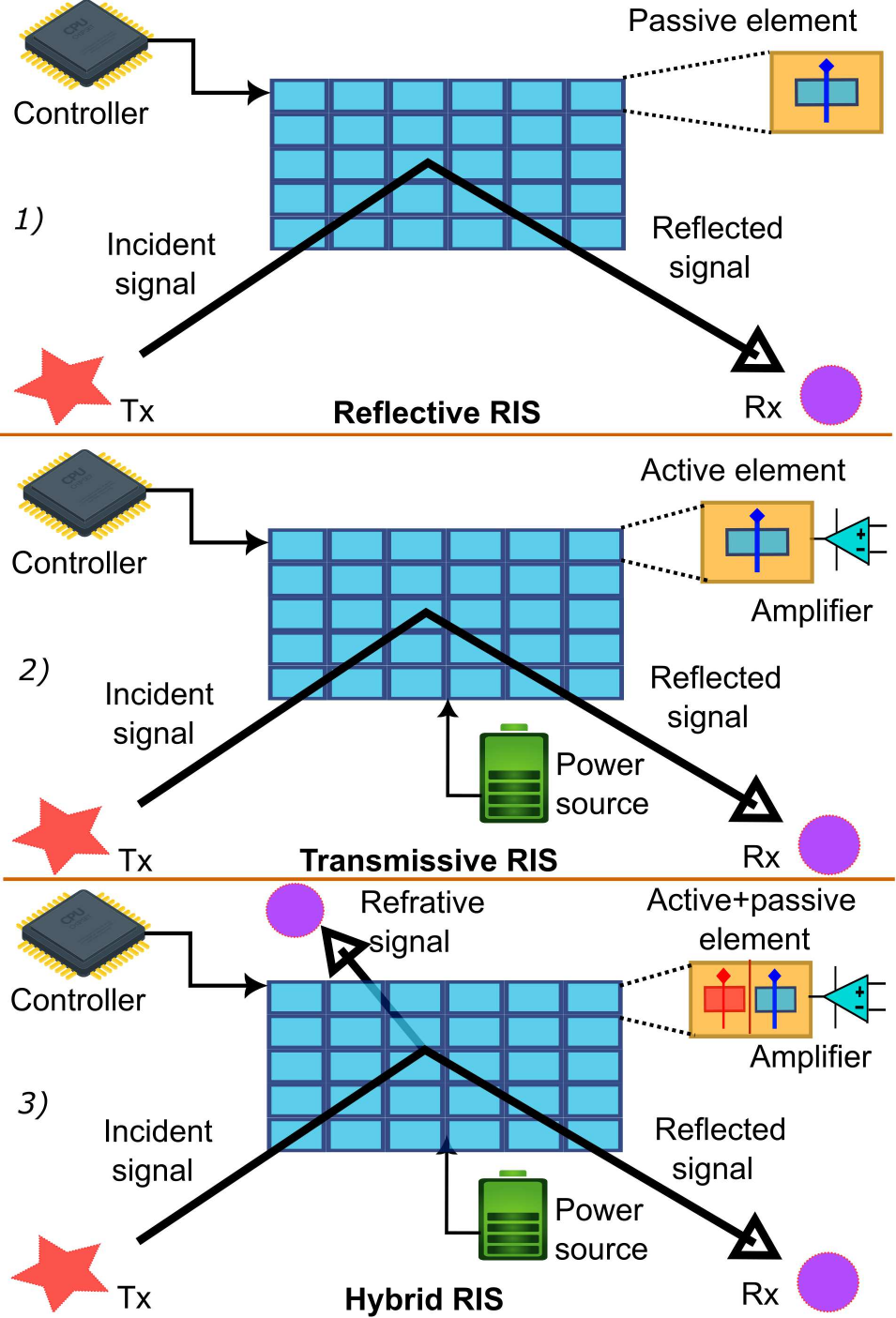}
\caption{RIS types based on operating mode.}
\label{RISType}
\end{figure}

RIS-integrated NTNs are envisioned to bring a myriad of new opportunities into the realm of next-generation wireless systems. Fig. \ref{IoTmagN} illustrates the potential use of RIS in NTNs, covering a multitude of novel applications at different altitudes.. Recent studies have analyzed and shown its potential across different application domains. Recent results on the problem of power consumption minimization and energy efficiency optimization are investigated in \cite{9726800,10159025,9947334,10149096}. The achievable gains in terms of sum-rate maximization for RIS-integrated NTNs are available in \cite{9917323,10145087,9969148,9896808,10035943}.
Such systems also possess the inherent capacity to significantly enhance the security of wireless systems, and some preliminary results on the physical layer security for RIS-integrated NTNs are available in \cite{9343764,10003076,9897065}. Performance analysis for such systems has been recently presented in \cite{10093979,9865217,10136395}. Yet, amid these valuable contributions, it remains imperative to underscore the necessity for a holistic, long-term vision concerning next-generation RIS-integrated NTNs. Such a visionary outlook is poised to chart the way toward achieving energy-efficient global connectivity facilitated by the prowess of RIS. 

Driven by the considerations elucidated above, the primary objective of this manuscript is to engage in a comprehensive discourse regarding the unresolved intricacies that pervade the nascent realm of RIS-integrated NTNs. This emerging field inherently demands a sustained and long-term research commitment to realize its full potential. In the first part of this article, our aim is to provide insight into the fundamentals of RIS technology by distinguishing the different emerging architectures. Subsequently, we explore the various potentials that RIS can offer in the NTNs domain, thus stimulating further research on this exciting topic. Next, we discuss in detail the recent advancements on this topic in detail, previously briefly mentioned above  \cite{9726800,10159025,9947334,10149096,9917323,10145087,9969148,9896808,10035943, 9343764,10003076,9897065,10093979,9865217,10136395}. The manuscript continues with evaluating an important use case of LEO communication with an RIS mounted on a HAP and we provide preliminary, yet extremely important, insights into the expected energy efficiency as a function of the RIS meta-elements for the single connected and fully connected RIS-integrated NTNs. Finally, we thoroughly examine the existing challenges in this domain, which are crucial for shaping the future of RIS-integrated NTNs. These challenges play a pivotal role in paving the way toward the next generation of NTNs, where RIS integration takes center stage and paves the way for reaching the ultimate goal of establishing global connectivity in an energy efficient manner.

\emph{Paper Organization:} The rest of the paper is organized as follows. In Section \ref{sec_2}, we present the fundamentals of RIS technology and provide in-depth motivation for the necessity of RIS in NTNs. In Sections \ref{sec_3} and \ref{sec_4} we present the recent advancements and the use case study, respectively. Finally, Sections \ref{sec_5} and \ref{sec_6} discuss the open challenges and the conclusions, respectively.


\section{RIS Fundamentals and Potential in NTNs}

 In this section, our aim is to outline various classifications of RIS. Subsequently, we delve into an exploration of the reasons underpinning the incorporation of RIS into NTNs.

\subsection{RIS Types Based on Operating Mode} \label{sec_2}
Based on operating mode, RIS systems can be classified into three commonly recognized types as illustrated in Fig. \ref{RISType}, 1) reflective RIS systems, 2) transmissive RIS systems, and 3) hybrid RIS systems.

{\bf Reflective RIS Systems}
The reflective RIS system consists of passive reconfigurable elements, such as metallic patches/antennas, which reflect/manipulate or redirect incident electromagnetic waves. The reflection or scattering properties of reflective RIS systems can be controlled by adjusting the physical dimensions and geometry of passive elements. A RIS system operating in reflecting mode does not require a power supply or active controller, making it simple, energy-efficient and cost-effective.

{\bf Transmissive RIS Systems}
The transmissive RIS system consists of active reconfigurable structural elements, such as transistors/phase shifters. The elements of active RIS allow for dynamic control over the reflection and transmission properties of the surface. By actively manipulating the electrical or magnetic characteristics of the RIS elements, a transmissive RIS system can provide more advanced functionalities, such as active/passive beamforming, phase control, and adaptive signal processing.

{\bf Hybrid RIS Systems}
The IRS system operating in hybrid mode consists of passive and active elements to enhance the system's performance and flexibility. Such RIS systems typically consist of a combination of fixed passive and controllable active elements. The passive elements of hybrid RIS provide a base reflection or scattering behaviour, while its active elements offer additional control and optimization capabilities. A hybrid RIS system aims to balance complexity, designing cost, and system performance.

\subsection{RIS Types Based on Architecture}
As described in Introduction Section, RIS can be categorized into three types: single connected, fully connected and group connected architectures. Let ${\bf \Phi}$ is the phase shift matrix of RIS having $M$ elements. Then, the non-zero elements in ${\bf \Phi}$ depends on the architecture of RIS. In the following, we have discussed three main architectures.

{\bf Single-connected IRS} 
In this architecture of IRS, the elements are not interconnected, and the phase-shift response is diagonal and can be expressed as ${\bf \Phi}=\text{diag}\{\varphi_1,\varphi_2,\dots,\varphi_M\}$, s.t. $|\varphi_m|^2=1,\ \forall m$.

{\bf Fully-connected IRS}
In the fully connected irs, elements are fully interconnected through reconfigurable impedance components and the phase shift response is the full matrix such that ${\bf \Phi}^H{\bf \Phi}=\boldsymbol{\delta}_M$, where $\boldsymbol{\delta}_M$ denotes the identity matrix.

{\bf Group-connected}
In the group connected IRS, elements are divided into $U$ groups, with each group being fully interconnected. For the convenience, each group consists of same number of element, i.e., $\Bar{M}=M/U$, where ${\bf \Phi}$ is a diagonal matrix which is given as ${\bf \Phi}=\text{bdiag}\{{\bf \Phi}_1,{\bf \Phi}_2,\dots,{\bf \Phi}_U\}$. Here ${\bf \Phi}_u\in\mathbb C^{\Bar{M}\times\Bar{M}}$ should satisfy ${\bf \Phi}^H_u{\bf \Phi}_u=\boldsymbol{\delta}_{\Bar{M}}$. 
\begin{table}
\centering
\caption{RIS architectures \& non-zero elements in $\boldsymbol{\Phi}$.}
\label{tab:1}
\resizebox{\columnwidth}{!}{%
\begin{tabular}{cccccl|}
\cline{2-6}
\multicolumn{1}{c|}{} &
  \multicolumn{1}{c|}{No. G} &
  \multicolumn{1}{c|}{GD} &
  \multicolumn{1}{c|}{E/G} &
  \multicolumn{1}{c|}{No. NZE} &
  Constraints \\ \hline
\multicolumn{1}{|c|}{SC} &
  \multicolumn{1}{c|}{$M$} &
  \multicolumn{1}{c|}{$1$} &
  \multicolumn{1}{c|}{$1$} &
  \multicolumn{1}{c|}{$M$} &
  $|\phi_m|^2=1, \forall m\in \mathcal{M}$ \\ \hline
\multicolumn{1}{|c|}{FC} &
  \multicolumn{1}{c|}{1} &
  \multicolumn{1}{c|}{$M$} &
  \multicolumn{1}{c|}{$M^2$} &
  \multicolumn{1}{c|}{$M^2$} &
  $\boldsymbol{\Phi}^H\boldsymbol{\Phi}=\boldsymbol{\delta}_M.$ \\ \hline
\multicolumn{1}{|c|}{GC} &
  \multicolumn{1}{c|}{$U$} &
  \multicolumn{1}{c|}{$M$} &
  \multicolumn{1}{c|}{$\Bar{M}^2$} &
  \multicolumn{1}{c|}{$U\Bar{M}^2$} &
  $\boldsymbol{\Phi}_u^H\boldsymbol{\Phi}_u=\boldsymbol{\delta}_{\Bar{M}}, \quad \forall u \in \mathcal{U}.$ \\ \hline
\multicolumn{6}{|l|}{\begin{tabular}[c]{@{}l@{}}SC: Single Connected, FC, Fully Connected, GC, Group Connected, \\ GD: Group Dimension, E/G: Element per Group, NZE; Non-Zero Elements\end{tabular}} \\ \hline
\vspace{-5mm}
\end{tabular}%
} 
\end{table}

\subsection{Why RIS in NTNs?}
RIS offers the following features in NTNs.

{\bf Improving Link Budget}
In NTNs, closing the communications link from the satellite to the ground terminal can be difficult because mobile ground users typically employ low-directional antennas, especially when operating at a high frequency. RIS in NTNs improves the link budget by compensating for path losses and improving signal reception at ground stations or user devices. This results in better data rates and increased system performance.

{\bf Enhancing Signal and Focusing}
Communications links between different platforms in NTNs, such as UAVs, HAPs, or satellites in NTNs, can be affected by path loss, signal attenuation, fading, and directional challenges. RIS can be deployed on these platforms to enhance the strength and directionality of signals. By intelligently reconfiguring signals toward desired locations, RIS enables improved signal focusing, resulting in better link quality and extended coverage.

{\bf Mitigating Interference and Reusing Spectrum}
The NTNs share same spectrum resources with terrestrial networks, resulting in co-channel interference caused to each network. Interference between terrestrial and NTNs can be a concern that can significantly compromise both networks' services. RIS can act as an interference mitigation tool, creating nulls or constructive interference in specific directions to minimize interference and enable efficient spectrum reuse.

{\bf Energy Efficiency and Green Communications}
Energy efficiency is essential in NTNs to prolong mission duration, reduce costs, support sustainability, and ensure uninterrupted connectivity. RIS can play a crucial role in enhancing energy efficiency and promoting green communications in NTNs. As RIS can operate passively without requiring active transmissions, it can minimize energy consumption and reduces the required transmission power. IRS improves signal propagation by strategically reflecting and focusing signals, leading to extended battery life in NTNs such as UAV communications. 

{\bf Resilience and Redundancy}
The presence of both terrestrial and NTNs can introduce diverse challenges, such as potential link blockages due to obstacles or atmospheric conditions. Moreover, terrestrial networks may be susceptible to physical damage, power outages, or natural disasters, leading to communication breakdowns. IRS can provide an additional layer of resilience and redundancy and redirect signals to alternative communications paths, leveraging the capabilities of non-terrestrial links to ensure continuous communications.
\begin{figure*}[!t]
\centering
\includegraphics[width=0.8\textwidth]{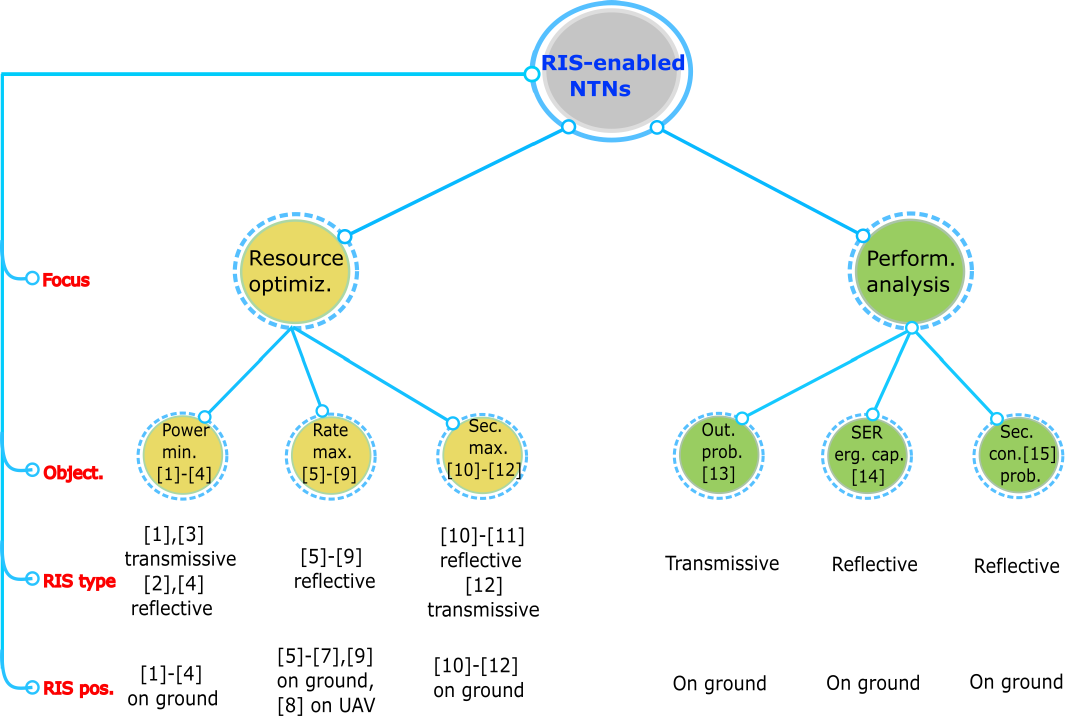}
\caption{Detailed of recent advances in RIS-integrated NTNs.}
\label{TaxIoTM}
\end{figure*}
\section{Recent Advances in RIS-integrated NTNs} \label{sec_3}
This section discusses and reports recent advances in NTNs involving RIS. The existing studies on RIS-integrated NTNs have focused on resource optimization and performance analysis. The authors of \cite{9726800,10159025,9947334,10149096} have carried out power consumption minimizations and energy efficiency solutions in IRS-enabled NTNs. In \cite{9726800}, Zhi {\em et al.} have minimized the transmit power of RIS-integrated hybrid terrestrial NTNs. They performed joint beamforming for satellite and base station (BS) and phase shift designs for RIS using alternating optimization, Taylor expansion and penalty function. Similarly, Bai {\em et al.} \cite{10159025} have investigated a power consumption minimization problem in RIS-integrated cognitive radio terrestrial NTNs. They adopted angular discretization for passive beamforming at RIS and successive convex approximation methods for active beamforming at satellite and BS. Moreover, Ge {\em et al.} \cite{9947334} have proposed an active beamforming scheme using alternating optimization to minimize the transmit power of BS while maximizing the secrecy rate of satellite users. Liao {\em et al.} \cite{10149096} have minimized power consumption in terrestrial NTNs following the Stackelberg game model. Using alternating optimization, the authors proposed hybrid beamforming for satellite, UAV, and RIS.

Researchers have also investigated sum rate maximization problems in RIS-integrated NTNs \cite{9917323,10145087,9969148,9896808,10035943}. In \cite{9917323}, Dond {\em et al.} have maximized the weighted sum rate of RIS-integrated NTNs. They jointly optimized the transmit beamforming, phase shift design and user scheduling using agglomerative hierarchical clustering and block coordinate descent methods. The same authors in \cite{10145087} have adopted deep reinforcement learning for optimizing active beamforming and phase shift design. They employed anti-jamming communications strategy to maximize the weighted sum rate of RIS-integrated NTNs. Moreover, Bai {\em et al.} \cite{9969148} have jointly optimized beamforming, phase shift design and power allocation to maximized the weighted sum rate of RIS-integrated NTNs. They employed successive convex approximation based on  second order Taylor expansion, Charnes-Cooper approach, and Bernstein-type inequality. Liu {\em et al.} \cite{9896808} have proposed an efficient optimization approach to maximize the ergodic sum rate of RIS-integrated NTNs. They use alternating optimization for power allocation, phase shift design and beamforming based on successive convex approximation, Charnes-Cooper and semi-definite programming methods. Furthermore, Song {\em et al.} \cite{10035943} have maximized the minimum covert rate in RIS-integrated NTNs by optimization the transmit precoding and RIS phase shift design. 

In addition, some researchers have studied physical layer security in RIS-integrated NTNs \cite{9343764,10003076,9897065}. The authors of \cite{9343764} have proposed a secure cooperative communications in RIS-integrated NTNs. They used alternating optimization approach for transmit beamforming and phase shift design optimization to minimize the received signals strength at eavesdropper. Khan {\em et al.} \cite{10003076} have proposed an alternating optimization strategy in RIS-integrated NTNs under multiple eavesdroppers scenario. They maximized the achievable secrecy capacity of the system by optimizing the transmit power and phase shift design. Moreover, the work in \cite{9897065} have exploited alternating optimization method for achievable secrecy rate maximization in RIS-integrated NTNs by optimizing beamforming, artificial noise, and phase shift design. Besides the resource optimization, some researchers have also investigated performance analysis of RIS-integrated NTNs \cite{10093979,9865217,10136395}. Guo {\em et al.} \cite{10093979} have derived an exact expression of outage probability in RIS-integrated NTNs. Another work \cite{9865217} has investigated closed-form expression of average symbol error rate and ergodic capacity in RIS-integrated NTNs. Off late, Ngo {\em et al.} \cite{10136395} have derived a closed-form expression of secrecy and connection probability in RIS-integrated Cache-aided NTNs.

\section{Case Study: RIS-integrated NTNs: Classical or Beyond Diagonal Architechtur?} \label{sec_4}
\begin{figure}[!t]
\centering
\includegraphics[width=0.47\textwidth]{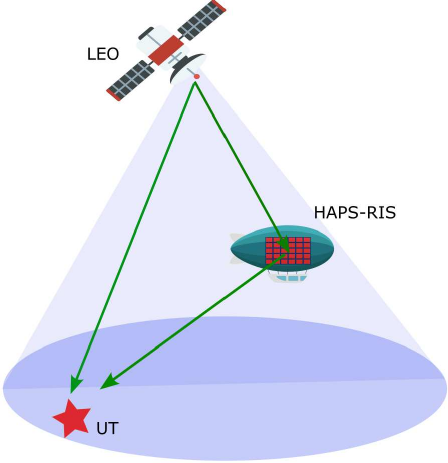}
\caption{System model.}
\label{IoTmagSM}
\end{figure}
As illusterated in Fig. \ref{IoTmagSM}, we consider a downlink LEO satellite communications, which provides services to a ground user terminal (UT). In satellite communications, closing the communications link from LEO to UT is difficult because UT typically employs low-directional antennas, mainly when operating at a high frequency. We consider a RIS system which is mounted on HAPS to assist the signal delivery from LEO to UT to address this issue and improve the link budget. Thus, UT receives a signal through a direct link and HAPS-RIS link. The RIS system consists of $K$ elements, while the transmitter and receiver use a single antenna scenario. This framework considers reflective mode of RIS with signal connected (classical RIS) architecture and fully connected (beyond diagonal RIS) architecture. It is assumed that the HAPS-RIS placement has already been optimized prior to RIS phase shift design. It is also assumed that the channel state information is available. 

This framework seeks to maximize the achievable energy efficiency of the system by enhancing the received channel gain of UT. This is achieved through efficient phase shift design at the HAPS-RIS system. The optimization of achievable energy efficiency maximization is formulated as non-linear concave/convex problem subject to the phase shift of RIS elements. In the proposed problem formulation, the energy efficiency is defined as the ratio of achievable rate of UT to the transmit power of LEO satellite. Based on the nature of the problem, a convex optimization toolbox of MATLAB is used to obtain an efficient solution.
\begin{figure}[!t]
\centering
\includegraphics[width=0.45\textwidth]{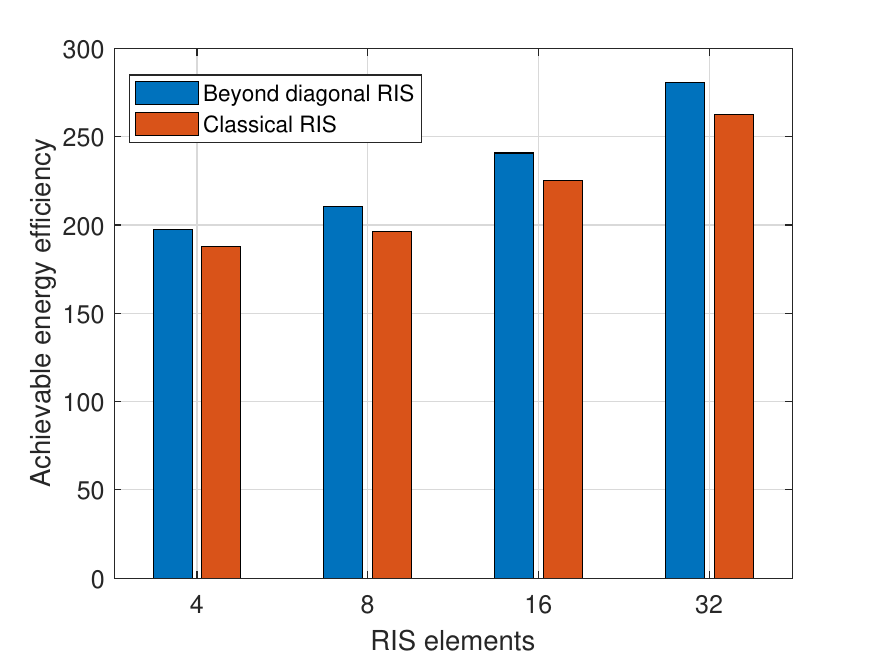}
\caption{Achievable energy efficiency versus number of RIS elements.}
\label{IoTMag}
\end{figure}

For the simulation setup, we set the carrier frequency as Ka-band (17.7-19.7 GHz), the transmit power of the satellite is 50 dBm, the bandwidth of UT is 20 MHz, the noise power density is -170 dBm, the number of RIS elements is 32, and the height of HAPS-RIS is 15 km. To assess system performance, we provide and compare results of fully connected RIS system (denoted as Beyond diagonal RIS) and single connected RIS system (sated as classical RIS). In classical RIS, all the elements independently manipulate the phase shift of the incident signal and constitute a diagonal matrix. In beyond diagonal RIS, elements are connected and cooperate with each other to manipulate the phase shift of the incident signal, and all matrices are full. Fig. \ref{IoTMag} shows the achievable energy efficiency versus the number of RIS elements. It can be seen that the system's achievable energy efficiency increases as the number of elements increases for both frameworks. This is because more RIS elements help further enhance the received channel gain without consuming any extra energy, which results in improving achievable energy efficiency. It can also be observed that the beyond diagonal architecture of RIS achieves more energy efficiency than the classical architecture of RIS. This is due to the cooperative nature and full phase shift matrix of beyond diagonal RIS. 

\section{Open Challenges} \label{sec_5}
In this section, we discuss and highlight current challenges and future research directions.

{\bf Hardware Complexity and Calibration}
Designing and fabricating RIS with precise elements can be technically challenging and costly, especially for large-scale platform deployment in NTNs. Moreover, large-scale deployments also require calibration and synchronization among the RIS elements to achieve coherent signal manipulation. Ensuring accurate and real-time control of RIS to optimize signal paths can be complex, especially when multiple platforms in NTNs are involved. The challenges of interoperability, standardization, and interfaces to enable the practical adoption of RIS in diverse NTN deployments must be addressed.

{\bf Dynamic Channel Conditions}
Different platforms in NTNs, especially those involving UAV and satellite communications, experience dynamic and time-varying channel conditions due to mobility and atmospheric effects. Conversely, RIS configuration and optimization depend on the channel state information acquisition and are crucial for seamless connectivity. Therefore, RIS must adapt to changing channel conditions in real-time to maintain optimal signal focusing and interference mitigation. Further, developing efficient algorithms for real-time channel estimation and control in dynamic NTN environments are required.

{\bf Artificial Intelligence/Machine Learning (AI/ML)}
AI/ML offers exciting opportunities to enhance RIS performance in NTNs. By utilizing AI/ML techniques and algorithms, RIS can optimize real-time signal reflection patterns, adapt to changing network conditions, predict channel variations, and self-optimize based on feedback. AI/ML-driven RIS can adjust its reflective properties dynamically, ensuring the best possible signal strength and quality, even in dynamic NTN environments. Integrating AI/ML into RIS promises more efficient, reliable, and adaptive communications in NTNs.

{\bf Security and Privacy}
As RIS manipulates signal propagation, ensuring the security and privacy of communications becomes essential. NTNs can be involved in sensitive applications, and protecting against potential signal manipulation or unauthorized access is of utmost importance. Researchers must focus on developing secure signalling protocols, robust authentication mechanisms, and privacy-preserving techniques to safeguard against unauthorized access and signal manipulation. Implementing signal integrity verification, intrusion detection, and access control measures can further enhance the security of RIS communications in NTNs.

\section{Conclusion} \label{sec_6}
The integration of RIS with NTNs has the potential to address several challenges, including enhancing received signal capacity in challenging environments, mitigating propagation losses in the atmosphere and space, overcoming link interruptions due to blockages, optimizing signal paths to counter dynamic channel conditions, improving energy efficiency by reducing the need for active transmissions, and providing secure and private communication links. This paper has provided the fundamental concept of RIS, reported recent advances in RIS-integrated NTNs, proposed a new framework for HAPS-RIS-integrated LEO satellite communications, and highlighted future research directions.

\ifCLASSOPTIONcaptionsoff
  \newpage
\fi

\bibliographystyle{IEEEtran}
\bibliography{Wali_EE}

\section*{Biographies}\small
\noindent {\bf Wali Ullah Khan [M]} (waliullah.khan@uni.lu) received a Ph.D.
degree in information and communication engineering from
Shandong University, Qingdao, China, in 2020. He is currently
working with the SIGCOM Research Group, SnT, University of Luxembourg.
\vspace{0.2cm}

\noindent {\bf Asad Mahmood [S]} (asad.mahmood@uni.lu) received
his Master degrees in Electrical Engineering from COMSATS University Islamabad, Wah Campus, Pakistan. He is currently pursuing the Ph.D.
degree with the Interdisciplinary Centre for Security,
Reliability, and Trust (SnT), University of Luxembourg.
\vspace{0.2cm}

\noindent {\bf Chandan Kumar Sheemar} (chandankumar.sheemar@uni.lu) received his Bachelor's and Master's degrees from the University of Padova, Italy, in 2016 and 2018, respectively, and his PhD from EURECOM, Sophia Antipolis, France, in 2022. He is currently a research associate in the SIGCOM Research Group, SnT, University of Luxembourg, Luxembourg. 
\vspace{0.2cm}

\noindent {\bf Eva Lagunas [SM]} (eva.lagunas@uni.lu) received a Ph.D. degree
in telecommunications engineering from the Polytechnic University of Catalonia (UPC), Barcelona, Spain, in 2014. She currently holds a research scientist position in the SIGCOM Research
Group, SnT, University of Luxembourg.
\vspace{0.2cm}

\noindent {\bf Symeon Chatzinotas [F]} (symeon.chatzinotas@uni.lu) received
Ph.D. degrees in electronic engineering from the University of Surrey, Guildford, United Kingdom, in 2009. He is currently a full professor or Chief Scientist I and the co-head of the SIGCOM Research Group, SnT, University of Luxembourg.
\vspace{0.2cm}

\noindent {\bf Bj\"orn Ottersten [F]} (bjorn.ottersten@uni.lu) received his Ph.D.
degree in electrical engineering from Stanford University, California, in 1990. He is currently the director for SnT, University of Luxembourg.

\end{document}